\documentclass{PoS}

\usepackage{multirow}
\usepackage{longtable} 
\usepackage{mathrsfs}  
\usepackage{textcomp}
\usepackage{natbib}
\usepackage{graphicx}
\usepackage{subcaption}
\usepackage{soul}

\def\civ{C{\sc iv}$\lambda1549$}
\def\kms{\,km\,s$^{-1}$}

\def\hb{{\sc{H}}$\beta$\/}
\def\mgii{Mg {\sc{ii}}}

\def\mbh{$M\mathrm{_{BH}}$}
\def\fblr{$f\mathrm{_{BLR}}$}

\def\DRhb{$\Delta R_{\mathrm{H\beta}}$}

\def\fvar{$F\mathrm{_{var}}$}
\def\mdotc{$\dot{\mathscr{M}}\mathrm{^{c}}$}
\def\RL{$R\mathrm{_{H\beta}}-L_{5100}$}

\def\fblrc{$f\mathrm{_{BLR}^{\,c}}$}
\def\DL{$D{\mathrm{_L}}$}
\def\RHb{$R\mathrm{_{H\beta}}$}

\title{Quasars from \textit{LSST} as dark energy tracers: first steps\footnote{on behalf of the \textit{LSST} AGN SC Collaboration}}

\ShortTitle{Quasars from \textit{LSST} as dark energy tracers}

\author{\speaker{Mary Loli Mart\'inez--Aldama}\thanks{Acknowledgements. The project was partially supported by National Science Centre, Poland, grant No. 2017/26/A/ST9/00756 (Maestro 9) and MNiSW grant DIR/WK/2018/12.}\\
        Center for Theoretical Physics, Polish Academy of Sciences\\
        E-mail: \email{mmary@cft.edu.pl}}

\author{Swayamtrupta Panda\\
        Center for Theoretical Physics, Polish Academy of Sciences\\
        Nicolaus Copernicus Astronomical Center, Polish Academy of Sciences\\
        panda@cft.edu.pl}
        
\author{Bo{\.z}ena Czerny\\
        Center for Theoretical Physics, Polish Academy of Sciences\\
        bcz@cft.edu.pl}

\author{Michal Zaja\v{c}ek\\
        Center for Theoretical Physics, Polish Academy of Sciences\\
        zajacek@cft.edu.pl}
        
\abstract{In the near future, new surveys promise a significant increase in the number of quasars (QSO) at large redshifts. This will help to constrain the dark energy models using quasars. The Large Synoptic Survey Telescope (\textit{LSST}) will cover over 10 million QSO in six photometric bands during its 10-year run. QSO will be monitored and subsequently analyzed using the photometric reverberation mapping (RM) technique. In low--redshift quasars, the combination of reverberation--mapped and spectroscopic results have provided important progress. However, there are still some facts which have to be taken into account for future results. It has been found that super-Eddington sources show time delays shorter than the expected from the well-known Radius-Luminosity (\RL) relation. Using a sample of 117 \hb\ reverberation--mapped AGN with $0.02<z<0.9$, we propose a correction by the accretion rate effect recovering the classical \RL\ relation. We determined the cosmological constants, which are in agreement with $\Lambda$--Cold Dark Matter model within 2$\sigma$ confidence level, which is still not suitable for testing possible departures from the standard model. Upcoming \textit{LSST} data will decrease the uncertainties in the dark energy determination using reverberation--mapped sources, particularly at high redshifts. We show the first steps in the modeling of the expected light curves for \hb\ and \mgii.}

\FullConference{Multifrequency Behaviour of High Energy Cosmic Sources - XIII - MULTIF2019\\
		3-8 June 2019\\
		Palermo, Italy}

\begin{document}

\section{Introduction}

Understanding the behavior of dark energy is one of the most challenging problems in physics and astrophysics nowadays. Dark energy is responsible for the accelerated Universe expansion, indicated by precise measurements based on Supernovae Ia, Cosmic Microwave Background (CMB), Baryon Acoustic Oscillations (BAO) and weak lensing. The most recent values of the cosmological constants: $\Omega_m=0.3111\pm0.0056$,   $\Omega_\Lambda$=0.6889$\pm$0.0056 and $Ho=$67.66$\pm0.42$ \kms\ Mpc$^{-1}$ \citep{planck2018},  suggest a flat geometry which satisfies the $\Lambda$--cold dark matter model ($\Lambda$CDM). Cosmological estimations are done using two observable indicators: standard rules and standard candles. Standard rulers are sources with a known angular size, which is converted to a physical size and it is then possible to determine the distance to the source. This is the technique used on the BAO estimations. On the other hand, standard candles are objects where the intrinsic luminosity is known and the luminosity distance (\DL) can be inferred. A classical example is Supernovae Ia (SN Ia), which have constrained cosmological models up to $z\sim1.4$. Higher redshift SN Ia are rare, and their use can be evolutionary biased.

The necessity of sources with larger redshift ranges for testing the cosmological models will be provided by the next generation of surveys. The \textit{{Large Synoptic Survey Telescope (LSST)}\footnote{Also known as the Vera Rubin Survey Telescope, for more details see \href{https://www.lsst.org/}{https://www.lsst.org/}}} will observe over 10 million quasars in six photometric bands during its 10-year run and analyzed using the photometric reverberation mapping technique. Reverberation mapping is based on the time delay between the emission line and the continuum variations.  This technique has been applied to around 100 AGN and quasars mainly in the \hb\ region, although there are some measurements using \civ\ and \mgii\ (e.g. \citet{grier2019, czerny2019cts}). The low number of reverberation-mapped sources is mainly due to the required long-term monitoring time, which will be solved by \textit{LSST}.

Reverberation mapping has provided important results on AGN and quasars physics, which are summarized in Section~\ref{sec:RL}. This technique has potential use in cosmology since it allows to determine the distance to the source in an independent way, applying the relation between the size of the broad line region (\RHb) and the monochromatic luminosity ($L_\lambda$). However, it has recently been found that high Eddington sources show a departure from \RL\ relation in the optical range (Section~\ref{sec:departure}) and are associated with the smallest continuum variations (Section~\ref{sec:variability}). We  proposed a correction by the accretion rate effect (Section~\ref{sec:departure}), which permit us to estimate the cosmological constants within 2$\sigma$ confidence level (Section~\ref{sec:cosmo}). The large uncertainties would be due to the low number of sources and the different techniques employed in the literature to estimate the time delay between the emission line and the continuum. This correction proposed by us needs to be taken into account for the future \textit{LSST} results. In Section~\ref{sec:lsst}  we present the first steps in modeling of the light curves expected for \hb\ and \mgii.

\section{Radius--Luminosity relation}
\label{sec:RL}

It has been observed that broad emission lines respond to the continuum variations with a time delay ($\tau \mathrm{_{obs}}$) of days or weeks. This time is directly related to the light travel across the broad-line region (BLR), i.e., $\tau \mathrm{_{obs}}$=~\RHb~/~$c$, where \RHb\ is the size of the BLR and $c$ is the speed of light. Reverberation mapping (RM) is based on the monitoring of the source to get $\tau \mathrm{_{obs}}$. It has provided information about the stratification of the BLR. For example, the time delay shown by high--ionization emission lines (HILs), like \civ, is smaller than that shown by low--ionization lines (LILs), like \hb. It means that HILs are emitted in a closer zone to the central continuum source. 

The most important result of RM is the Radius--Luminosity relation. In the optical range \citep{bentz2013}, it is given by:
 \begin{equation}
\mathrm{log}\left(\frac{R\mathrm{_{H\beta}}}{\mathrm{1 lt-day}}\right)= (1.527\,\pm\,0.31)\, +\, 0.533{^{+0.035}_{-0.033}}\, \mathrm{log}\left(\frac{ L_{5100}}{10^{44}L_\odot}\right).
\label{equ:bentz}
\end{equation}
By knowing the luminosity of the source, we can determine the size of the BLR and thus other properties like the black hole mass and the accretion rate. 

\begin{figure*} 
\centering
\includegraphics[width=0.51\textwidth]{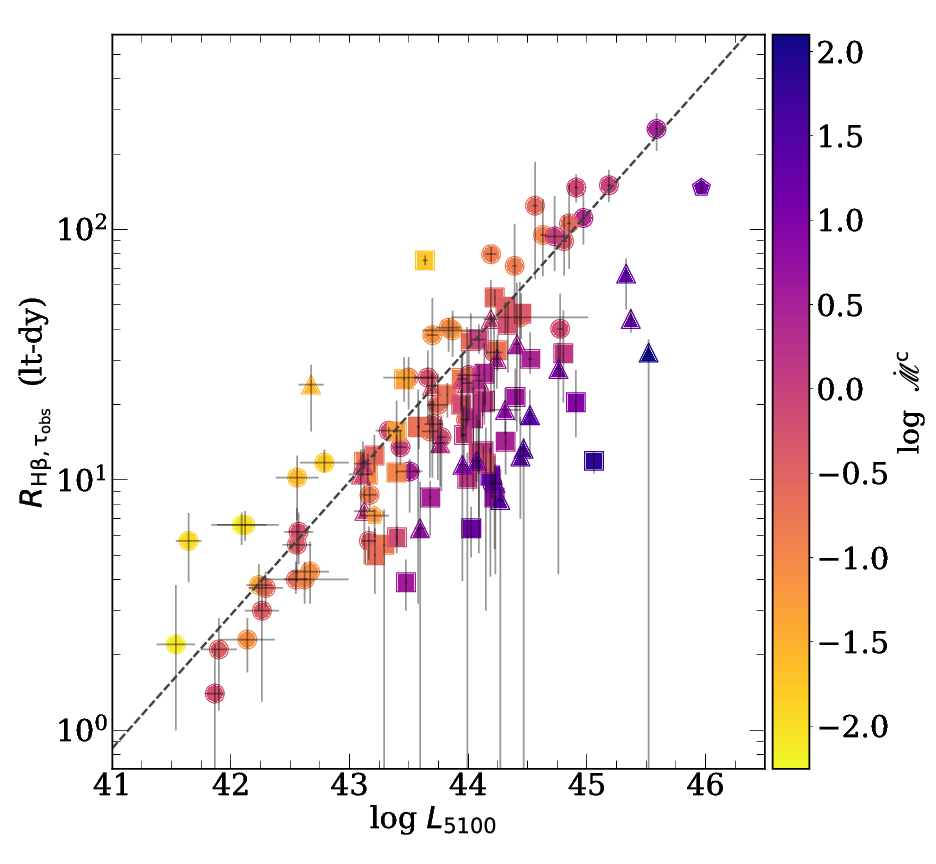}
\includegraphics[width=0.48\textwidth]{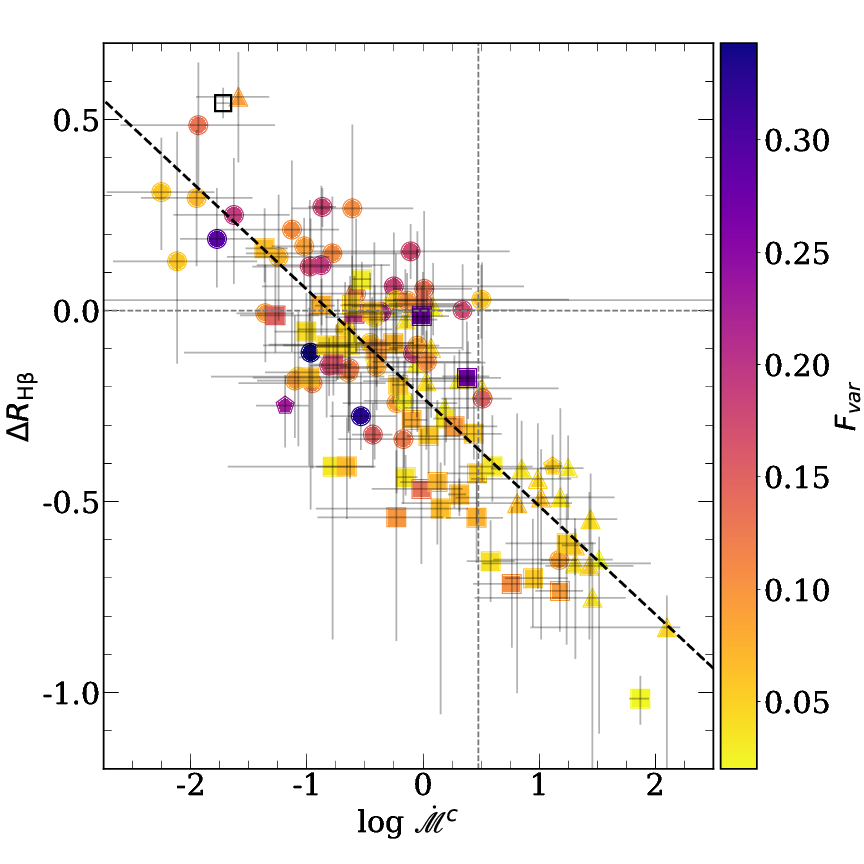}
\caption{LEFT PANEL: $R_{H\beta}-L_{5100}$ relation for SEAMBH (triangles), SDSS-RM (squares), Bentz Collection (circles), NGC5548 and 3C 273 (pentagons). Colors indicate the variation in dimensionless accretion rate, \mdotc. Dashed black line corresponds to the expected $R_{H\beta}-L_{5100}$ relation from \citet{bentz2013} (Equation~\ref{equ:bentz}). RIGHT PANEL: Relation between \DRhb\ and \mdotc. Marker colors indicate the variation in \fvar\ at 5100\AA. In both panels, the parameters have been estimated with an anti-correlated virial factor, \fblrc. SEAMBH stands for super-Eddington accreting massive black holes \citep{wang2014}, the SDSS-RM sample is taken from \citet{grier2017} and the Bentz Collection is described in Section 2 in \citet{martinezaldama2019}. \label{fig:RL-DR}}
\end{figure*}

Using a sample of 117 \hb\ reverberation-mapped sources, we built a \RL\ diagram, which is shown in the left panel of Figure~{\ref{fig:RL-DR}}. To estimate the black hole mass (\mbh), we considered the virial factor anti-correlated with the full-width at half maximum (FWHM) of the emission line (\fblr) proposed by \citet{mejia-restrepo2018}, which includes a correction for the orientation effect. Accretion rate is estimated considering the dimensionless accretion rate (\mdotc) described by \citet{du2015}. The assumptions in the estimation of these parameters could include some bias in our analysis, which are discussed in the Section~\ref{sec:caveats}. \mbh\ and \mdotc\ values, and details of the sample are reported in \citet{martinezaldama2019}. 

The left panel of Figure \ref{fig:RL-DR} illustrates the variation of the dimensionless accretion rate through  \RL\ diagram. Sources with the largest \mdotc\ values show the largest departures from the \RL\ relation, i.e., their time delays are smaller than that predicted by the \RL\ relation \citep{du2015,du2018}. Super-Eddington sources show a particular behavior compared with the rest of the AGN. \citet{wangshielding2014} found that their continuum is produced by a slim disk. Their BLR tends to show large densities ($n_{H}\sim10^{13}$  cm$^{-3}$) and low-ionization parameters (log $U<-2$) \citep{negrete2012}. Also, some spectroscopic features are characteristic of this kind of sources, like the strong intensity of very low-ionization lines (e.g. optical FeII) or the blue asymmetries in the HIL profiles \citep{martinezaldama2018}. All these features are directly related to the high accretion rate shown by these sources, which is also responsible of the departure for the \RL\ relation.  


\section{Departure from the \RL\ relation}
\label{sec:departure}
 
\begin{figure*} 
\centering
\includegraphics[width=0.55\textwidth]{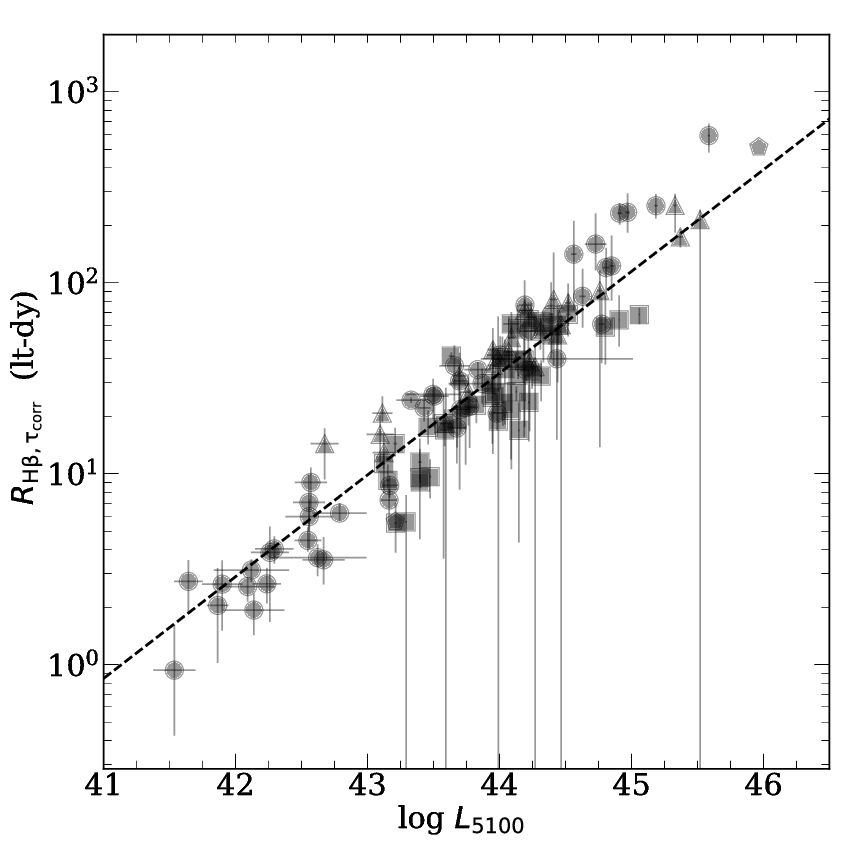}
\caption{\RL\ relation corrected by the dimensionless accretion rate effect. Symbols are the same as Figure \ref{fig:RL-DR}. \label{fig:RLcorr}} 
\end{figure*}

The departure from the \RL\ relation is estimated by the parameter \DRhb\ defined as $\Delta R_{\mathrm{H\beta}}=\mathrm{log}\,\left( \frac{\tau\mathrm{_{obs}}}{\tau\mathrm{_{{H\beta_{\,R-L}}}}}\right)$, where $\tau\mathrm{_{{H\beta_{\,R-L}}}}$ is the time delay expected from the \RL\ relation and can be estimated from the Equation~\ref{equ:bentz}. In right panel of Figure~{\ref{fig:RL-DR}} is shown the behavior of \DRhb\ as a function of the \mdotc. There is a strong relation between these two parameters, which is supported by the Pearson coefficient ($P=0.822$) and  $rms$ (0.172) values. In order to describe this relation, we performed an orthogonal linear fit, which is given by: 
\begin{equation}\label{equ:DRHb}
\Delta R\mathrm{_{H\beta, \dot{\mathscr{M}}\mathrm{^{c}}}} = \left(-0.283\pm0.017\right) \mathrm{log}\dot{\mathscr{M}}^\mathrm{{c}}+\left(-0.228\pm0.016\right)
\end{equation}
Therefore, the largest accretion rates correspond to the largest departures. With this relation we can correct the observed time delay using the relation:
\begin{equation} \label{equ:taucorr}
\tau_{\mathrm{corr}}(\dot{\mathscr{M}}^{c})\,=\,10^{-\Delta R\mathrm{_{H\beta}( \dot{\mathscr{M}}\mathrm{^{c}}})} \, \cdot \tau_{\mathrm{obs}}.
\end{equation}
With this correction, we recover the low scatter ($\sigma\mathrm{_{corr}}=0.396$) along the \RL\ relation as is shown in the Figure~\ref{fig:RLcorr}. 

\subsection{Remarks on \mbh\ and \mdotc\ estimations}
\label{sec:caveats}

Black hole mass is estimated from the relation: $M\mathrm{_{BH}}=f\mathrm{_{BLR}}\, \frac{R\mathrm{_{H\beta}} \,\, FWHM^2}{G}$, where \fblr\ is the virial factor which includes information about the dynamics, structure and orientation of the BLR. Many formalisms have been proposed to get a general scheme of its behavior \citep[e.g.][and references therein]{onken2004,collin2006,woo2015,yu2019}, however, the large diversity of AGN properties complicate the scenario. The virial factor uncertainty affects directly the black hole mass determination introducing an error by a factor 2--3.

The proper way to calibrate the virial factor is by the comparison with an independent method to get the black hole mass. Typically, the relation between the \mbh\ and the bulge or spheroid stellar velocity dispersion ($\sigma_*$), the relation \mbh--$\sigma_*$  \citep[e.g.][]{2000ApJ...539L...9F,gultekin2009,woo2015} is used. Although the estimation is based on the $rms$ spectrum of the reverberation--mapped AGN, which is not $strongly$ affected by orientation effects, there are still large uncertainties when it is applied to the general AGN population. Also, there is a lack of super-Eddington sources in the estimations.

\citet{collin2006} found that the virial factor changes according to the shape of the profile (Gaussian or Lorentzian). Sources with narrower profiles would be associated with virial factors larger than broad profiles. The viral factor proposed by \citet{mejia-restrepo2018} shows this behavior, \fblrc=$\left(\frac{{\mathrm{FWHM_{obs}}}}{4550 {\pm 1000}}\right)^{ -1.17}$ (expression for the \hb\ emission line). Recently, an independent analysis done by \citet{yu2019} provided a similar expression for the virial factor, which supports the selection for our analysis. 

However, \fblrc\ has some caveats. It has been performed using a sample with a predominance of broad profiles (FWHM$>3000$ \kms), then the results could change with the inclusion of a large number of sources with narrower profiles. One--third of our sample has narrow profiles (FWHM$<2000$ \kms), which also corresponds to the super-Eddington sources. This fact would change \mdotc\ values and the correction proposed by the accretion rate effect (Equation~\ref{equ:DRHb}). 

The selection of a variable virial factor recovers in some sense the orientation effect, according to \citet{strochi-bergmann2017} (see also \citealt{panda2019b}).  
We repeated the analysis using a fixed virial factor, \fblr=1, which is typically used in the single-back hole mass estimations. The scatter shown in the relation between \DRhb\ and the dimensionless accretion rate is larger than that shown using a virial factor anti-correlated with the line FWHM, $rms=0.172$ vs. $rms=0.243$ respectively. Also, the relation is weaker according to Pearson coefficient value ($P=0.572$). Therefore, it suggests that a variable virial factor corrects the orientation effect.

\section{Variability}
\label{sec:variability}

The colors in the right panel of Figure \ref{fig:RL-DR} show the behavior of \fvar, which estimates the $rms$ of the intrinsic variability relative to the mean flux \citep{rodriguez-pascual1997}. \fvar\ mimics the behavior of the amplitude of variability. In the Figure an anti-correlation between \fvar\ and the dimensionless accretion rate is observed. The Spearman coefficient ($\rho {_{s}}=-0.374$ with a probability of $p=1.6\times10^{-5}$) indicates a weak relation. A similar correlation has been found by several authors in larger samples \citep[e.g.][and references therein]{sanchez-saenz2018}. \textit{LSST} will be able to measure the variability properties like \fvar, which can be used as a tool in the identification of the physical properties of the AGNs. 

\section{Hubble diagram}
\label{sec:cosmo}

\begin{figure*} 
\centering
\includegraphics[width=0.49\textwidth]{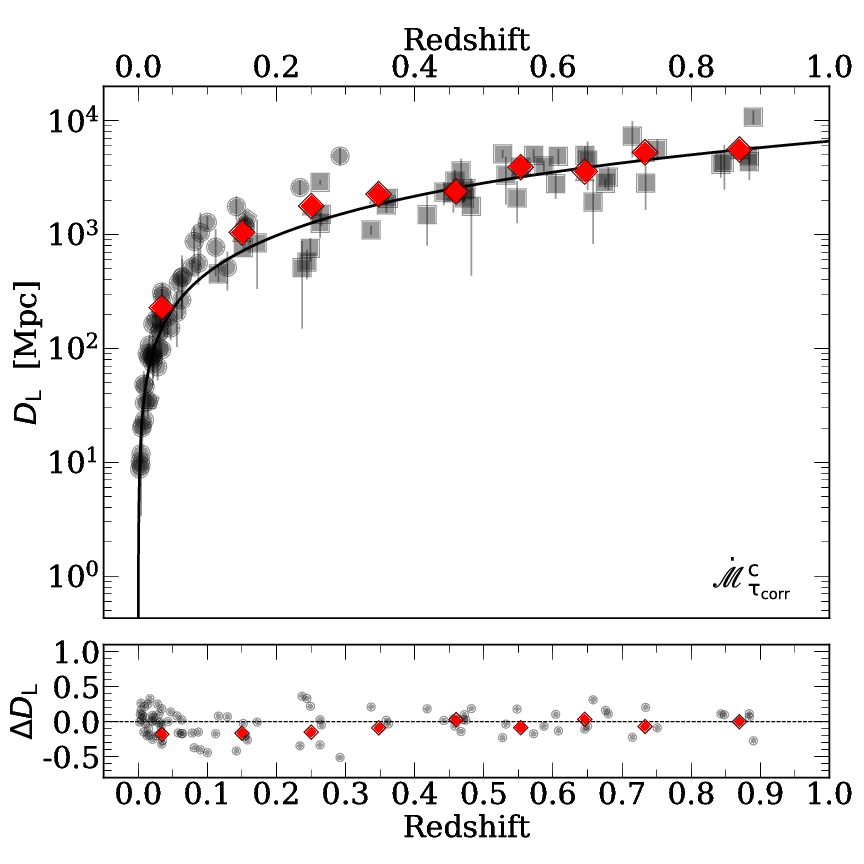}
\includegraphics[width=0.47\textwidth]{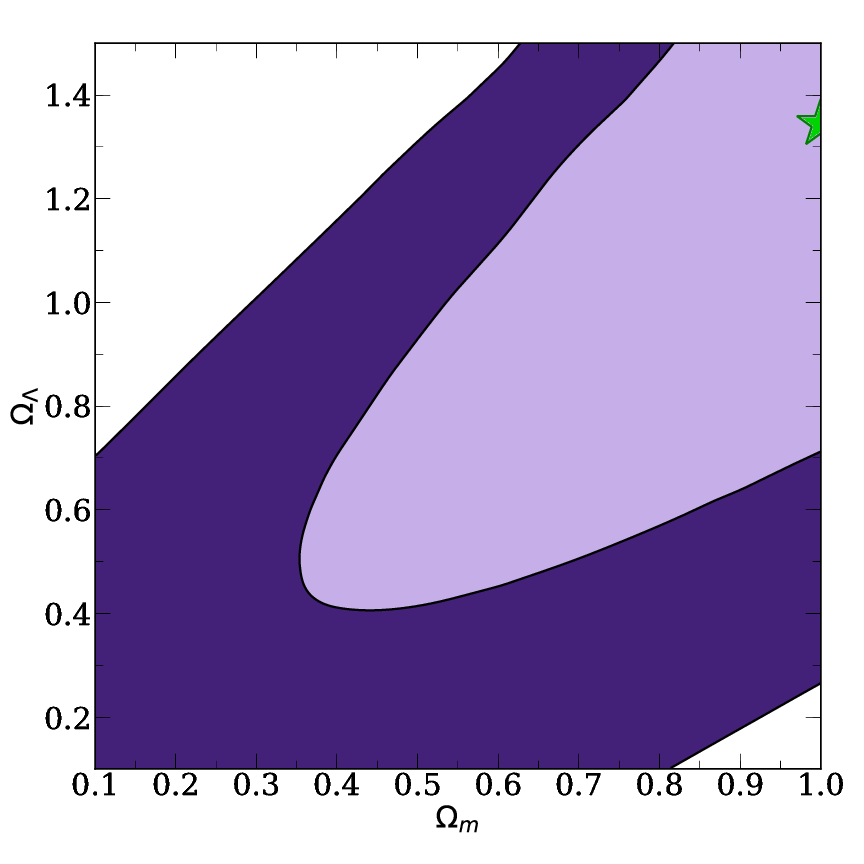}
\caption{LEFT PANEL: Hubble diagram obtained from the time delay corrected by the dimensionless accretion rate effect. Gray symbols are the same as Fig. \ref{fig:RL-DR}. Red symbols correspond to the mean of the luminosity distance in redshift bins of $\Delta z = 0.1$. The black line indicates the expected luminosity distance based on the standard $\Lambda$CDM model. The bottom panel shows the difference between the expected luminosity distance and the observed one. RIGHT PANEL: $\chi^2$ behavior in the $\Omega_m-\Omega_\Lambda$ space for the full sample. Contours correspond to 1$\sigma$ and 2$\sigma$ confidence levels, respectively.  Green symbol marks the minimum $\chi^2$ value. \label{fig:HD-contour}}
\end{figure*}

Using the corrected time delay  by the accretion rate effect ($\tau_\mathrm{_{corr}}$) and the luminosity at 5100\AA\ ($L_{5100}$), we estimate the luminosity distance using the equation $D\mathrm{_L}= \left(\frac{ L_{5100}}{ 4\,\pi\,F_{5100} }\right)^{1/2}$. Then, we built a Hubble diagram, which is shown in the left panel of Figure~\ref{fig:HD-contour}. In the diagram, we also show the luminosity distance expected from the classical $\Lambda$CDM model with parameters: $H_0=67$ km s$^{-1}$ Mpc$^{-1}$, $\Omega_{\Lambda}=0.68$, $\Omega_m=0.32$ (black line). We also mark the luminosity distance for redshift bins of $\Delta z=0.1$ (red symbols), these values are only included for visualization, since the number of points are not enough for statistical analysis. 

In order to get a determination of the cosmological parameters, we estimate the best cosmological model. We assumed a standard $\Lambda$CDM model, and the value of the Hubble constant, $H_0=67$ km s$^{-1}$ Mpc$^{-1}$, and search for the minimum value (best fitting) using the function:
\begin{equation}
    \chi^2 = \sum^{N}_{i=1} \frac{ (\mathrm{log}(D_\mathrm{L,mod}^{\,i}) - \mathrm{log}(D_\mathrm{L,obs}^{\,i}))^2} {(\mathrm{log}(1 + b_ i)^2 + \sigma^2)},
\end{equation}
where $N$ is the total number of sources in the sample, $b_i$ is the relative error in the luminosity distance determination and $\sigma$ is the dispersion in the sample  described by \citet{risaliti2015}. 

The best fit is shown in the right panel of Figure \ref{fig:HD-contour} (green symbol), which is consistent within 2$\sigma$ confidence level with the standard $\Lambda$CDM model. This result indicates that errors are still large, they are probably related to the small number of sources and the different methods employed to determine the time delay. Using different time-delay determination methods (interpolated cross-correlation function-ICCF, discrete cross-correlation function - DCF, $z$-transformed DCF, JAVELIN, randomness estimators-Von Neumann), we showed that differences arise when applied to the same data, in particular uneven and heterogeneous datasets, specifically concerning  the time-delay uncertainties and even the different peaks of the time-delay distribution \citep{2019arXiv190703910Z}.

\section{Prospect for  \textit{LSST} }
\label{sec:lsst}

\textit{LSST} \citep{lsst2019} is a 8.4m telescope with a \textit{state-of-the-art} 3.2 Gigapixel flat-focal array camera that will allow to perform rapid scan of the sky with 15 seconds exposure and thus providing a moving array of color images of objects that change. Every night, \textit{LSST} will monitor $\sim$75 million AGNs and is estimated to detect $\sim$300+ million AGNs in the $\sim$18000 deg$^2$ main-survey area \citep{Luo2017}.

\textit{LSST} is a photometric project but the 6-channel photometry can be effectively used for the purpose of reverberation mapping and estimation of time delays. We present some preliminary results from our code in development which allows to produce \textit{mock} light curves and recover the time delays. The code takes into consideration several key parameters to produce these light curves, namely -- (1) the campaign duration of the instrument (10 years); (2) number of visits per photometric band; (3) the photometric accuracy (0.01-0.1 mag)\footnote{these values are adopted from \citet{lsst2019}}; (4) black hole mass distribution\footnote{the results shown here are for a representative black hole mass, M$_{\rm{BH}}$ = 10$^8$ M$_{\odot}$.}; (5) luminosity distribution\footnote{the results shown here are for two representative cases of bolometric luminosity, L$_{\rm{bol}}$ = 10$^{45}$ and 10$^{46}$ erg s$^{-1}$.}; (6) redshift distribution\footnote{the results shown here are for two representative cases of redshifts, z = 0.1 and 0.985.}; and (7) line equivalent widths (EWs) consistent with SDSS quasar catalogue \citep{shen11}. We create continuum stochastic lightcurve for a quasar of an assumed magnitude and redshift from AGN power spectrum with Timmer-Koenig algorithm \citep{tk1995}. The code takes as an input a first estimate for the time delay measurement. We utilize the standard \RL\ relation \citep{bentz2013} to estimate this value. In the current version of the code, the results for the photometric reverberation method are estimated by adopting only 2 photometric channels at a time and the time delay is estimated using the $\chi^2$ method. We account for the contamination in the emission line (H$\beta$, MgII) as well as the in the continuum. The code also incorporates the FeII pseudo-continuum and contamination from starlight i.e. stellar contribution.  

\begin{figure*}
        \centering
        \begin{subfigure}[b]{0.475\textwidth}
            \centering
            \includegraphics[width=\textwidth]{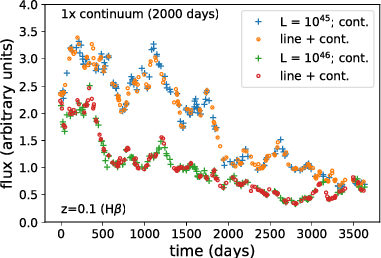}
        \end{subfigure}
        \hfill
        \begin{subfigure}[b]{0.475\textwidth}  
            \centering 
            \includegraphics[width=\textwidth]{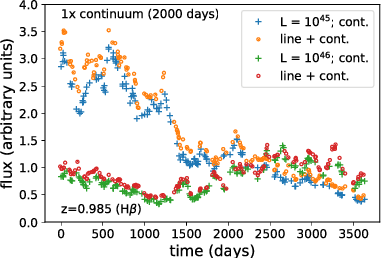}
        \end{subfigure}
        \vskip\baselineskip
        \begin{subfigure}[b]{0.475\textwidth}   
            \centering 
            \includegraphics[width=\textwidth]{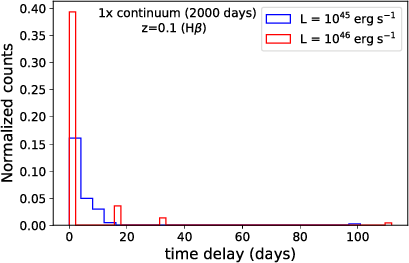}
        \end{subfigure}
        \quad
        \begin{subfigure}[b]{0.475\textwidth}   
            \centering 
            \includegraphics[width=\textwidth]{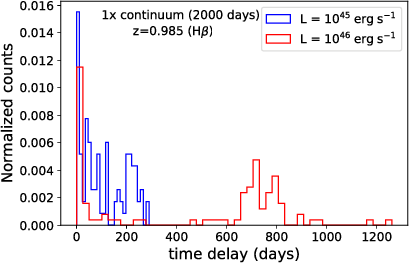}
        \end{subfigure}
        \caption{Preliminary lightcurves for H$\beta$ at \textit{z} = 0.1 (UPPER-LEFT) and \textit{z} = 0.985 (UPPER-RIGHT). The lightcurves are generated using \textit{Timmer-Koenig} method \citep{tk1995} and convolved with photometric data distribution expected from \textit{LSST} \citep{lsst2019} with arbitrary normalization. Two cases of luminosity ($10^{45}$ and $10^{46}$ erg s$^{-1}$) are shown, each for the continuum and the line. The time axis refers to the monitoring time (in days) in the upper panels. The corresponding time delay distributions are shown in the lower panels.}
        \label{fig:lsst01}
    \end{figure*}

\begin{figure*}
        \centering
        \begin{subfigure}[b]{0.475\textwidth}
            \centering
            \includegraphics[width=\textwidth]{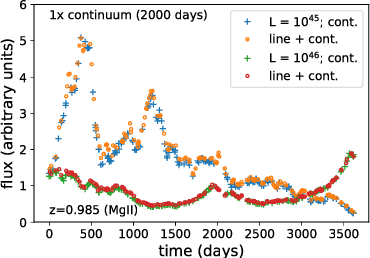}
        \end{subfigure}
        \hfill
        \begin{subfigure}[b]{0.475\textwidth}  
            \centering 
            \includegraphics[width=\textwidth]{lightcurve_combine_Hb_0985.png}
        \end{subfigure}
        \vskip\baselineskip
        \begin{subfigure}[b]{0.475\textwidth}   
            \centering 
            \includegraphics[width=\textwidth]{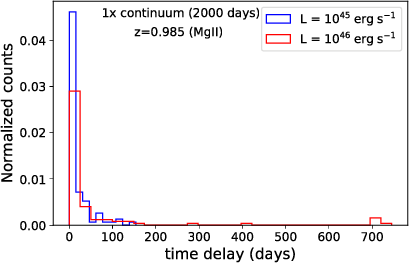}
        \end{subfigure}
        \quad
        \begin{subfigure}[b]{0.475\textwidth}   
            \centering 
            \includegraphics[width=\textwidth]{delay_distribution_normed_Hb_0985.png}
        \end{subfigure}
        \caption{Preliminary lightcurves for MgII (UPPER-LEFT) and H$\beta$ (UPPER-RIGHT) at \textit{z} = 0.985, with arbitrary normalization. The corresponding time delay distributions are shown in the lower panels. The representation is same as Figure \ref{fig:lsst01}.}
        \label{fig:lsst02}
    \end{figure*}

{Since \hb\ and \mgii\ are the typical virial estimators at low and high redshift respectively, we performed a first test of their simulated lightcurves.} In the upper panels of each of Figures \ref{fig:lsst01} and \ref{fig:lsst02}, we show the variation in the simulated lightcurves for H$\beta$ as a function of the redshift and luminosity. Here, the time axis refers to the source monitoring time (in days). In the lower panels of each of Figures \ref{fig:lsst01} and \ref{fig:lsst02}, we show the corresponding time delay distributions. These time delays are obtained using the $\chi^2$ method, where, we have first assumed that the delay is null. We interpolate between the data points for the two channels and starting with an assumed shift between the two interpolated light curves to be null, we loop over the light curves to estimate the best shift over the distribution using cross-correlation. This best shift at each instance is represented in the form of time-delay histograms. In Figure \ref{fig:lsst02}, we show the variation in the lightcurves for two emission lines (MgII and H$\beta$) at the same redshift.  Similar to the Figure \ref{fig:lsst01}, we illustrate the lightcurves for two cases of luminosities.

The density of the photometric data-points is adopted from \citet{lsst2019}. An increase in the density of the points, both in the continuum channels and the line channels, leads to better recovery of the time delay measurements \citep{2019arXiv190905572P}, i.e., in the \textit{Deep-Drilling Fields} \citep{lsst_wp}. The power spectral distribution for these lightcurves is assumed to have the low-frequency break corresponding to 2000 days.


\section{Conclusions and work in progress}

{After 30 years of the beginning of the first reverberation mapping analysis, our knowledge of the AGN physics has progressed significantly. And, although $\sim100$ of sources have been monitored with this technique, their results have established the base for many other areas of AGN research. However, AGN with peculiar properties, like high accretion rate, have questioned their generalization of the \RL\ to the rest of AGN population. In this paper, we demonstrated that a correction can be applied to recover the expected \RL\ results. This correction is dependent on the chosen parameters, like the virial factor that  is still under debate,  but gives the first hints to understand and solve the problem. Upcoming \textit{LSST} data will also test all these results, giving a statistical confirmation or generating new questions. }

The code in development, although at a basic stage, is able to produce decent estimates of time-delay measurements. We are currently working to extend the analyses to include all six photometric channels simultaneously, and retrieve the time delay measurements to mimic actual observations. We have also tested the effect of having a denser coverage of photometric data points by scaling the continuum and the line contributions \citep{2019arXiv190905572P} where we are now able to estimate the time-delays with an accuracy within 1-3$\%$ error. We have also begun testing with different methodologies to generate lightcurves e.g. Damped-Random Walk and JAVELIN \citep{2011ApJ...735...80Z,2013ApJ...765..106Z,2016ApJ...819..122Z}. This also applies to the time-delay estimation where we are testing the consistency of the predicted values with other available methods e.g. Interpolated cross-correlation function \citep{1987ApJS...65....1G} and JAVELIN \citep{2011ApJ...735...80Z,2013ApJ...765..106Z,2016ApJ...819..122Z}. Another property that should be taken into consideration is the potential contribution of the jet and its components to the optical luminosity, which can affect the positioning of the source within \RL\, relation in terms of the luminosity. In particular, correlations have been found between the radio spectral slope, radio loudness, the luminosity of [OIII] line and the strength of FeII line \citep{2019A&A...630A..83Z,2019MNRAS.482.5513L}. More refined results will be presented in the near future.


\bibliographystyle{astron}
\bibliography{delays}


\bigskip
\bigskip
\noindent {\bf DISCUSSION}

\bigskip
\noindent {\bf JAMES BEALL:} What plan do you have for analyzing such volume of data?

\bigskip
\noindent {\bf MARY LOLI MART\'INEZ--ALDAMA:} {We will focus on sources with spectroscopic measurements. \textit{LSST} will be assisted by an array of campaigns \citep{brandt2017} during its proposed run. Moreover, there is a need for follow-up spectroscopic campaigns i.e. SDSS-V Black Hole Mapper \citep{SDSS5}, which aim to derive BLR properties and reliable SMBH masses for distant AGNs with expected observed-frame reverberation lags of 10--1000 days. The SDSS-V Black Hole Mapper survey will also perform reverberation mapping campaigns in three out of the four \textit{LSST} DDFs. These spectroscopic follow-up observations will allow us to choose the sources with \textit{better quality}. With our code, we intend to prepare not only a mock catalog that will mimic the real observations that will be done by \textit{LSST}, but also, use the real data to correct them that will then be user-ready for subsequent analyses. For more details about the data pipelines using \textit{LSST} see \citet{lsst2019}.}

\end{document}